\def\year{2023}\relax
\documentclass[letterpaper]{article} 
\usepackage{aaai23}  
\usepackage{times}  
\usepackage{helvet} 
\usepackage{courier}  
\usepackage[hyphens]{url}  
\usepackage{graphicx} 
\usepackage{natbib}  
\usepackage{caption} 
\usepackage{type1cm} 
\usepackage{graphicx} 
\usepackage{xspace} 
\usepackage{balance} 
\usepackage{booktabs} 
\usepackage{multirow} 
\usepackage[tableposition=top]{caption} 
\usepackage{bold-extra} 
\usepackage{siunitx} 
\usepackage[vlined,linesnumbered,ruled,noend]{algorithm2e} 
\usepackage{microtype} 
\usepackage{xfrac} 
\usepackage{mathtools} 
\usepackage[capitalise]{cleveref}
\usepackage{subfig}
\usepackage[dvipsnames]{xcolor}
\graphicspath{{../img/}}

\newenvironment{squishlist}
{\begin{list}{$\bullet$}
 {\setlength{\itemsep}{0pt}
	\setlength{\parsep}{3pt}
	\setlength{\topsep}{3pt}
	\setlength{\partopsep}{0pt}
	\setlength{\leftmargin}{1.5em}
	\setlength{\labelwidth}{1em}
	\setlength{\labelsep}{0.5em} } }
{\end{list}}

\title{Authority without Care:\\Moral Values behind the Mask Mandate Response}

\author{
    Yelena Mejova,\textsuperscript{\rm 1}
    Kyriaki Kalimeri,\textsuperscript{\rm 1}
    Gianmarco De Francisci Morales\textsuperscript{\rm 2}\\
}
\affiliations{
    \textsuperscript{\rm 1}ISI Foundation, V. Chisola 5, Turin, Italy,
    \textsuperscript{\rm 2}CENTAI, C. Inghilterra 3, Turin, Italy\\
    yelenamejova@acm.org, kyriaki.kalimeri@isi.it, gdfm@acm.org

}
\begin{document}

\maketitle

\begin{abstract}
Face masks are one of the cheapest and most effective non-pharmaceutical interventions available against airborne diseases such as COVID-19.
Unfortunately, they have been met with resistance by a substantial fraction of the populace, especially in the U.S.
In this study, we uncover the latent moral values that underpin the response to the mask mandate, and paint them against the country's political backdrop.
We monitor the discussion about masks on Twitter, which involves almost 600k users in a time span of 7 months.
By using a combination of graph mining, natural language processing, topic modeling, content analysis, and time series analysis, we characterize the responses to the mask mandate of both those in favor and against them.
We base our analysis on the theoretical frameworks of Moral Foundation Theory and Hofstede's cultural dimensions.

Our results show that, while the anti-mask stance is associated with a conservative political leaning, the moral values expressed by its adherents diverge from the ones typically used by conservatives.
In particular, the expected emphasis on the values of authority and purity is accompanied by an atypical dearth of in-group loyalty.
We find that after the mandate, both pro- and anti-mask sides decrease their emphasis on care about others, and increase their attention on authority and fairness, further politicizing the issue.
In addition, the mask mandate reverses the expression of Individualism-Collectivism between the two sides, with an increase of individualism in the anti-mask narrative, and a decrease in the pro-mask one.
We argue that monitoring the dynamics of moral positioning is crucial for designing effective public health campaigns that are sensitive to the underlying values of the target audience.
\end{abstract}

\section{Introduction}

The COVID-19 pandemic has been an unprecedented event, which has also brought about an infodemic that makes public health response difficult.
COVID denial, anti-vaccine sentiment, and other flavors of theories (from doubts to full-blown conspiracies) have been documented in social media \cite{patwa2021fighting}.
Among the many controversies, surprisingly, wearing a mask has become extremely politicized and contentious. 
Inconsistent messaging by public health organizations has seriously undermined public compliance to this simple measure.
Initially, the WHO recommended that masks should be worn only by professionals or those taking care of a sick person~\cite{who2020advice}, and
in the U.S., the Center for Disease Control and Prevention (CDC) recommended the use of masks only for healthcare workers and people who were sick.
Only on April 3rd, 2020, the CDC officially recommended wearing non-medical cloth face coverings when in public places~\cite{landsverk2020cdc}. 
Soon after, speculations that masks do more harm than good~\cite{gillespie2020mask} and that they foster a false sense of security~\cite{khan2020mask} started to proliferate.
Public health experts have acknowledged that it has been a challenge to communicate with groups holding strong values of freedom~\cite{hartsoe2020experts}.
Indeed, globally, research has found that the value of individualism is associated with higher COVID-19 mortality rates~\cite{maaravi2021tragedy} and less adherence to preventive measures~\cite{biddlestone2020cultural}.
However, how personal moral values evolve around governmental messaging concerning these preventive measures is not yet understood.


Mask-wearing behavior and attitudes can be influenced by multiple factors.
There is extensive evidence that political ideology and identity influences attitudes, judgments, and behaviors~\cite{van2018partisan}.
At the same time, political identity is deeply intertwined with moral values~\cite{graham2009liberals,hatemi2019ideology}.
Finally, and perhaps obviously, moral values are directly linked to moral decision making~\cite{karandikar2019predicting}, which completes the triad.
In particular, there is an altruistic component in mask-wearing, as their main purpose is to protect others, which is related to solidarity towards the in-group in the face of an out-group threat~\cite{campbell1965ethnocentric}.
For this reason, among moral values we pay particular attention to the individualist-collectivist angle.
Finally, there is ample literature on the division and polarization of news and information sources in general.\footnote{\url{https://www.pewresearch.org/journalism/2020/03/04/about-one-fifth-of-democrats-and-republicans-get-political-news-in-a-kind-of-media-bubble}}  
As users increasingly rely on social media to satisfy their information needs,\footnote{\url{https://www.pewresearch.org/journalism/2021/11/15/news-on-twitter-consumed-by-most-users-and-trusted-by-many}}  
the lines between professional reporting and personal opinions begin to blur.
For this reason, we contextualize the information environment in which the mask debate takes place in terms of information sources from peers---within the Twitter community---and from external sources.

This work provides a fine-grained analysis of the moral values of those expressing opinions around masking by applying the Moral Foundations Theory to a dataset of Twitter posts spanning the beginning of the pandemic, from January to July 2020. 
In particular, we ask:
\emph{What is the anatomy of the collective discussion on mask wearing around the mask mandate on Twitter?}
In particular, we analyze different facets of this discussion in the U.S.:
\begin{squishlist}
\item[1.] How does the users' stance relate to their political leaning?
\item[2.] What moral values do adherents to pro- or anti-masking stances hold?
\item[3.] What is the information environment around their arguments? 
\end{squishlist}

Our findings confirm the known political divisions, with liberals supporting the pro-mask and conservatives the anti-mask stance.
The moral narrative of the two sides also differs: those on the anti-mask side invoke the values of \emph{authority} and \emph{purity}, while those promoting mask-wearing emphasize \emph{care} and \emph{loyalty}.
The introduction of the mask mandate shifts these values from an emphasis on \emph{care} to an increased attention on \emph{authority} and \emph{fairness}, thus escalating the politicization of the issue.
Content analysis shows that the increasingly individualistic views espoused by the anti-mask side are accompanied by the allegations of the dangers and ineffectiveness of mask-wearing, supported by resources from social media and sometimes even governmental and scientific sources.
We argue that the mask-wearing debate has a dynamic moral landscape that should be carefully monitored to design effective public messaging campaigns that reflect the shifting values of their target audience.
Such studies are especially important, given the rising prominence of social media as a platform for public policy discussion and influence~\cite{shapiro2017politicians}. 





\section{Related Work}

\subsubsection{Masking Behavior.}
Mask mandates have been shown to be effective in decreasing COVID infection rates in the U.S., both in urbanized and rural areas \cite{krishnamachari2021role}. 
Early studies have tested the effect of mask adoption, with simulations showing that even relatively ineffective face coverings could meaningfully reduce community transmission~\cite{eikenberry2020mask}. 
Unfortunately, the public response to the masking mandates has been partial, with only a 12\% increase in mask-wearing immediately after the CDC masking guideline on April 3, 2020~\cite{goldberg2020mask}.
A survey of residents of 10 states in the U.S. relates mask-wearing to COVID cases in the state, political party in power, and individual measures of ``social capital'', as well as some demographics~\cite{hao2021understanding}. They find that respondents are more likely to wear a mask if there are more COVID-related deaths in the state (Odds Ratio of $1.26$), the Democratic party is in power (OR $=2.0$), and if the respondents more often speak to their friends and family (OR $=1.16$). 
Although misinformation has been blamed for the resistance to mask-wearing and social distancing, a nationally representative survey has found that the beliefs about the consequences of these behaviors are more predictive of people's compliance~\cite{hornik2021association}. 
Especially in the U.S., beliefs and trust in authority are often strongly related to the people's political affiliations.
For instance, Republicans, conservatives, and nationalists are less likely to believe that the World Health Organization (WHO) can effectively manage the pandemic \cite{bayram2021trusts}.




\subsubsection{Social Media \& Masking.} A study of tweets during the early days of the COVID pandemic (Feb-March 2020) identified methods for decreasing the spread of COVID as one of the main themes and the wearing of masks as one of those most associated with positive sentiment~\cite{abd2020top}. 
However, authors of a later study spanning January to October 2020 showed that the output of automated sentiment analysis tools corresponds poorly with the mask-related sentiment expressed in the tweets due to the richness of the language used \cite{he2021people}.
Instead, they perform manual coding of a sample of the tweets to identify several major categories of concerns around masking, including physical discomfort, effectiveness, appropriateness, and political beliefs. 
A network analysis of the mask-related tweets has shown the pro-mask activists exist in a kind of ``echo chamber''~\cite{cinelli2021echo,garimella2018political}, and that they tend to ignore the subversive rhetoric of the anti-mask fringe \cite{lang2021maskon}.
Instead, a recent Twitter study found a focus on ongoing news \cite{cotfas2021unmasking}.
This fringe is more likely to use toxic language, including insults and profanity, than the pro-mask ones~\cite{pascual2021toxicity}.
The authors link this behavior to either the vociferous protestations of a minority group \cite{miller2009expressing} or potential signaling as an in-group behavior and a marker of personal identification.
Alongside this toxicity, other studies find widespread misinformation and misunderstandings in the social media discussions around mask use. These include the beliefs that COVID19 is over-hyped by the media, that masks are ineffective, and that they do more harm than good \cite{keller2021social}. These beliefs were then shown to impact the mask-wearing (and social distancing) behaviors \cite{hornik2021association}.
In this study, we introduce a dimension of \emph{moral values}, which we argue underlies some of the disagreements on the appropriate use of masks during the epidemic, and which may shed some light on the moral reasoning behind the rhetoric. 


\subsubsection{Theoretical Framework.}
Our study of the rhetoric around masks during COVID-19 is grounded in the Moral Foundations Theory (MFT), its manifestation in interpersonal and inter-group communication, and its reflection at the societal level as an individualist or collectivist cultural dimension. 
According to social identity theory, members of an in-group will look for negative aspects of an out-group, thus enhancing their self-image~\cite{tajfel1979integrative}.
Strong in-group and out-group reasoning at a societal level determines whether a society is individualistic or collectivist (IC),
as defined by~\citet{hofstede2001culture}. 
The IC dimension considers the degree to which societies are integrated into groups and their perceived obligations and dependence on groups.
Individualism indicates there is a greater importance placed on attaining personal goals. 
Collectivism indicates there is a greater importance placed on the goals and well-being of the group. 
People in collectivist societies generally distinguish sharply between in- and out-groups, while people in individualistic societies treat everyone as a potential in-group member and thus apply universal values to everyone.
Cross-cultural research has demonstrated that the United States is the prototypical individualist culture based on the IC dimension
\cite{hofstede1984culture,kim1994individualism}.
At the same time, the U.S. show a measurable variation on this dimension~\cite{vandello1999patterns} at state level.

The societal dimensions of collectivism and individualism can be related to the individuals' adherence to specific moral dimensions, as postulated by the Moral Foundations Theory~\cite{Graham2012,samuel2016authoritariansim}, which include:
 \emph{care/harm}, fundamental concerns for the suffering of others,  including virtues of caring and compassion;
 \emph{fairness/cheating},  concerns about unfair treatment,  inequality,  and more abstract notions of justice;
 \emph{loyalty/betrayal},  concerns related to obligations of group membership;
 \emph{authority/subversion},  concerns related to social order and the obligations of hierarchical relationships such as obedience,  respect,  and proper role fulfillment; and
 \emph{purity/degradation}, with concerns about physical and spiritual contagion,  including virtues of chastity,  wholesomeness, and control of desires.
These foundations are shown to underlie human judgements and decision-making \cite{Weber2015} on societal topics ranging from vaccine hesitancy~\cite{kalimeri2019human,beiro2023moral}, to politics~\cite{iyer2012understanding}, religion, and social cooperation~\cite{haidt2012righteous, curry2016morality}.
Here, we place the focal point on the linguistic analysis of values expressed via Twitter, and aim to clarify peoples' dispositions and attitudes towards interpersonal and inter-group processes related to persuasion and communication narratives.

The moral dimensions of MFT have also been linked to political ideologies \cite{day2014shifting,kugler2014another}, with conservatives emphasizing the in-group relationships and tradition, and liberals endorsing fairness and equal opportunity. 
In the United States, the mask-wearing measure has also been strongly associated with the partisan divide. 
Social identity is a primary reason behind people's decision whether to wear a mask during the pandemic \cite{powdthavee2021face}, and surveys show that faith in President Trump is a strong predictor of refusal to social distance, and its effect is largest among individuals high in binding foundations
\cite{graham2020faith}. 
Indeed, the U.S. counties that showed strong support for Trump in 2016 practiced significantly lower mask-wearing in 2020 \cite{kahane2021politicizing}.
This work illustrates the strong connection between the attitudes expressed towards mask-wearing on Twitter and the the political leaning of those expressing them, and shows the shifts in moral emphasis after the mask mandates are introduced.


\section{Data}

We begin by collecting tweets mentioning the keywords ``mask'', ``facemask'', ``ffp3'', and ``n95'' (the latter two refer to popular kinds of masks), spanning the dates of January 1st to July 30th, 2020, using the GOT3 library~\cite{getoldtweets}.
These keywords were chosen by considering the special Twitter Covid-19, stream\footnote{\url{https://developer.twitter.com/en/docs/twitter-api/tweets/covid-19-stream/filtering-rules}} and picking the most common English keywords related to masks.
This collection results in \num{18245298} tweets from \num{5935103} users.
Following recommendations from existing literature, we then perform several filtering steps in order to ensure that the tweets can be used to assess the stance of the user on masking, and that the account is likely to belong to a human living in U.S.:

\begin{enumerate}
\item Relevance classifier (details next).
\item Exclude users whose location cannot be mapped to one of U.S. states or Washington DC.
\item Exclude those not having at least 1 English tweet \cite{garimella2016quantifying}.
\item Exclude users with only 1 tweet \cite{garimella2016quantifying}.
\item Exclude top 0.1\% of users by the number of tweets (having higher posting rate)~\cite{des2022detecting}.
\item Exclude users whose friends to followers ratio is $>$10~\cite{aggarwal2012phishari}.
\end{enumerate}

We proceed by making sure the tweets we collect are indeed about mask-wearing due to the COVID-19 epidemic.
Upon manual examination, we find several other topics captured, such as advice on masking during protests, sports-related wear, and beauty products.
To remove such content, we use a set of \emph{distant} labels to identify the non-relevant content.\footnote{We use lists compiled by finding top bigrams (considering the 300 most frequent) in the dataset and then manually labeling them for non-relevance. The irrelevant list is \{`hair mask',`gas mask',`sleeping mask', `majora's mask',`ski mask',`eye mask',`clay mask',`tear gas'\} and the relevant list is \{`covid',`coronavirus',`sars-cov-2',`sars-cov2',`social distanc',`socialdistanc'\}.}
We manually annotate a balanced random selection of 300 tweets associated with a distant label and find 96\% accuracy in distinguishing between relevant and non-relevant tweets (expected accuracy of a random baseline is 50\%).
We use \num{430568} tweets with the respective distant labels to train the relevance classifier.
The model is a logistic regression, trained on tf-idf-weighted unigram counts extracted from the tweets and with inverse-proportional class weighting to mitigate class imbalance.
As text pre-processing, we remove URLs, numbers, handles, hashtag (the \# symbols), retweet indicators, and stopwords, keep whole words, and perform snowball stemming.
The accuracy in 5-fold cross-validation is 99.7\%.
We draw another balanced sample of 300 tweets thus classified and find the accuracy of the classifier to be 82.3\% (inter-annotator agreement overlap of 90\%, Cohen's kappa of 0.76 between 3 annotators). 

Next, we geolocate the tweets identified as relevant by mapping the user location strings to the Geonames ID by using custom string matching,\footnote{Code available at \url{https://sites.google.com/site/yelenamejova/resources}} and to U.S. zip codes by using the `uszipcode' library.\footnote{\url{https://pypi.org/project/uszipcode/}}
We apply basic pre-processing (i.e., stopword and non-ASCII character removal).
First, we filter locations assigned to the U.S. territory by `uszipcode' and then the remaining ones by the Geonames library, aiming to recover the GPS coordinates of the smallest possible administrative area.
We are able to locate \num{1383729} of the users within a U.S. state or Washington DC.
Our sample is representative of the U.S. population distribution (2019 Census estimates) with a Pearson correlation of $0.96$ ($n$=51), which suggests that there is little bias in the sampling of the state in terms of number of users.

The latter four filtering steps aim at excluding users with either too little engagement on the topic or those who post so much that they are likely to be business or automated accounts.
We do not use the popular tool Botometer, as a recent study on mask-related tweets shows it to mostly find active human Twitter users \cite{he2021people}.
At the end of this process, we are left with \num{647730} users.
Finally, we use the Twitter API Friends call to collect the information about whom these users follow (``followees'' or ``friends''), thus resulting in the coverage of \num{598792} users.

\begin{table}[t]
    \caption{Number of tweets and associated users at different stages of data collection.} 
    \label{tab:dataset}
    \centering
    \begin{tabular}{l r r}
    \toprule
    Dataset stage & N tweets & N users \\
    \midrule
    Keyword-based collection & \num{18245298} & \num{5935103} \\
    Geo-location filter & \num{5685866} & \num{1383729} \\
    Engagement \& network filter & \num{3614343} & \num{598792} \\
    \bottomrule
\end{tabular}
\end{table}

\section{Results}

\subsubsection{Stance Classification.}
\begin{figure}[t]
\centering
\includegraphics[width=0.8\linewidth]{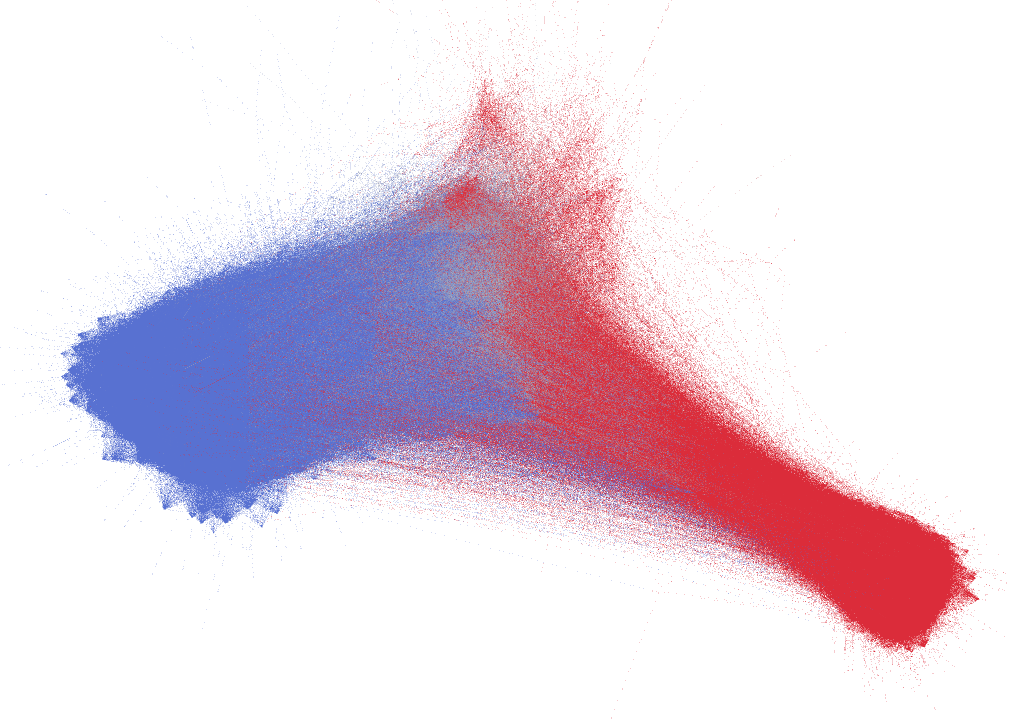} 
\caption{GCC of follower network, colored by METIS score.}
\label{fig:gcc_plot}
\end{figure}
Following previous work on identifying controversial topics on social media~\cite{garimella2016quantifying,garimella2018quantifying}, we look for a bi-partitioning of the network that would indicate polarization.
We look for two sides of the debate because, ultimately, the decision whether to wear a mask is binary.
We start our analysis by considering the follower network induced by the users captured in our data, shown in Figure~\ref{fig:gcc_plot}.
In total, we are able to retrieve \num{509755293} follower-followee relationships, such that followers are all users whose tweets relate to masks.
When we constrain the followees to the set of users in our tweet dataset, the network contains $|V| = \num{598792}$ users and $|E| = \num{35763336}$ edges.
We use the graph partitioning algorithm METIS~\cite{karypis1998fast} to partition the network into two groups, repeatedly $N=100$ times with different random seeds, so to get an ensemble of partition assignments for each node, and use the average partition assignment for each node across the $N$ repetitions as a polarity score $p \in [0,1]$~\cite{mejova2022modeling}.\footnote{Out of the two possible relabeling of each partition assignment, we choose the one that minimizes the distance from the current node labeling to maintain the partition identity fixed.}
We tune the relative size of the partitions by maximizing the number of users within \num{95}\% confidence interval of either extreme, and find the optimal proportion to be $1.1$:$1$ of pro-mask to anti-mask users.
As the algorithm does not indicate the actual stances of the users (in fact, it uses no content information whatsoever), each partition is assigned its stance manually by examining 10 sampled users on each side.
All users in the same partition expressed the same stance, which suggests a good separation.
The sampled users' stance is then propagated to all users within their partition, $0$ for pro-mask and $1$ for anti-mask.

\begin{figure}[t]
\centering
    \subfloat[Polarity score distribution]{\includegraphics[width=0.49\linewidth]{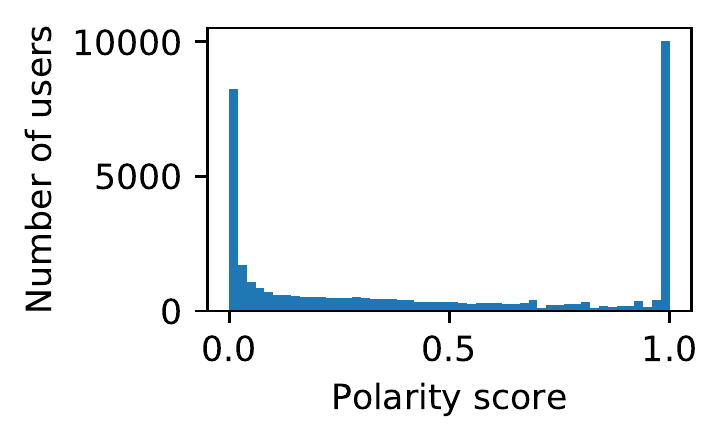}\label{fig:metis_hist_score}}
    \subfloat[Pol. leaning per mask stance]{\includegraphics[width=0.48\linewidth]{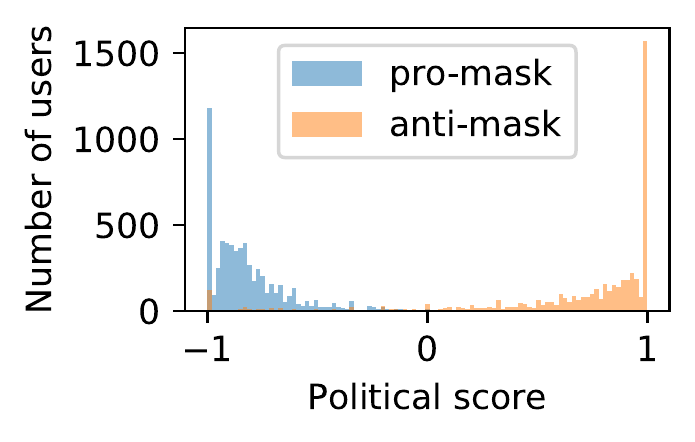}\label{fig:metis_hist_politics}}
    \caption{Distributions of (a) polarity scores of users (computed via METIS), and (b) political leaning of users (inferred via follower relationships on Twitter) grouped by inferred stance on mask-wearing.}
\label{fig:metis_hist}
\end{figure}

We also verify the robustness of our result by looking at the community structure of the network.
The top-4 communities found by the Louvain algorithm~\cite{blondel2008fast}, which include more than $90\%$ of the nodes of the network, have median polarity scores of $0.2, 1.0, 0.0, 1.0$, which indicate clear separation and polarization in their composition.
The largest community, which includes \num{18708} users and is mainly pro-mask, has a slightly larger spread of polarity scores, which may indicate a larger divergence of opinions (and possibly doubts) in the population.
This result can also be seen in Figure~\ref{fig:metis_hist_score}, the histogram of polarity scores of the network, which shows a smoother decline on the left side than on the right side.
Finally, we assign a discrete label to each node according to its polarity score.
We heuristically choose the threshold $0.05$, which allows us to assign a label to $56.4\%$ of users: $28.8\%$ with 0 and $27.6\%$ with 1, thus leaving $43.5\%$ of users with an unknown label.\footnote{The 95\% Agresti-Coull confidence interval upper bound for binomial proportions of a polarity score of $0$ is $0.0444$, which we round to $0.05$.}
The roughly equal shares of pro- and anti-mask users are in line with recent other manual annotation work on this topic \cite{cotfas2021unmasking}.

The two sides are colored as blue and red, and unknown as grey in Figure~\ref{fig:gcc_plot}.
To assess the accuracy of the approach described so far, we sample 60 users from those assigned a stance label by METIS and annotate them manually.
To determine the manual label, we use all captured tweets by each user, thus considering in aggregate \num{1866} tweets.
The annotation was done by three authors, all knowledgeable in the U.S. mask debate.
Twelve percent of the data was used for inter-annotator agreement, which shows perfect agreement.
The exercise revealed an overall precision of $86.4\%$ (95\% confidence interval [0.76, 0.94]), with perfect precision for pro-mask class, but only $72.4\%$ for anti-mask case, with several users incorrectly labeled as anti-mask by the algorithm.
The network structure suggests some connection of people who express doubts but are not clearly anti-mask with more extreme positions.
Indeed, it is likely that the anti-mask stance is more diverse than the pro-mask one, and cannot be easily represented with a single stance, similarly to the vaccine hesitancy case~\citep{cossard2020falling}.

\begin{figure}
\centering
    \includegraphics[width=.9\linewidth]{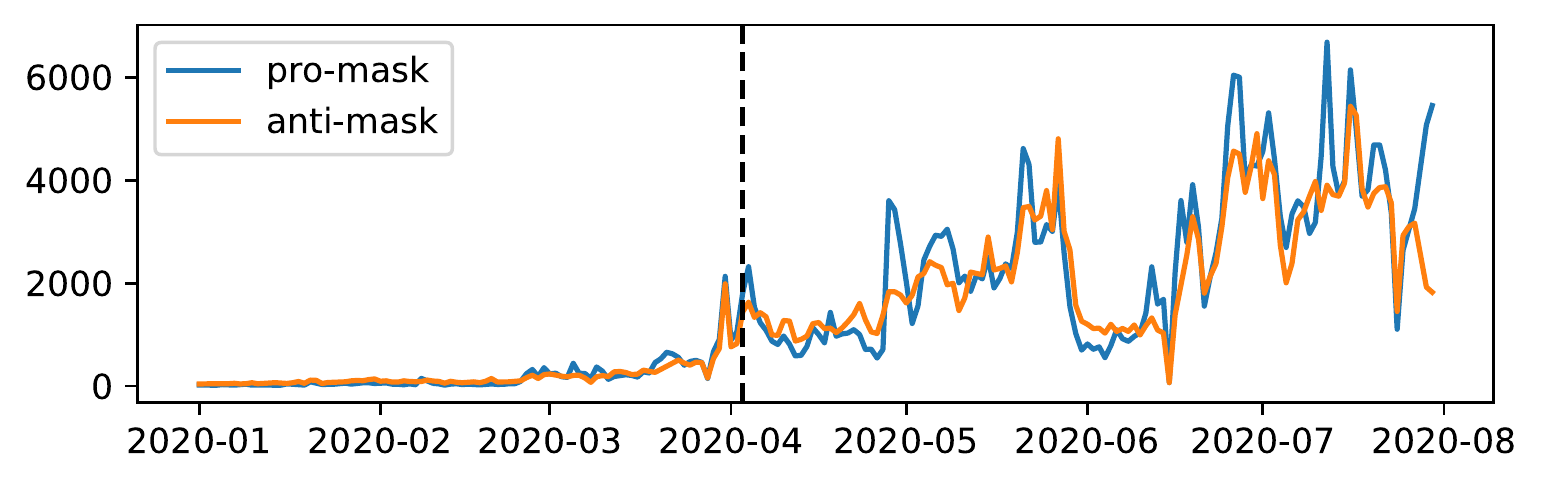}
    \caption{Daily time series of number of tweets by users classified by their mask stance. Vertical line on mandate date.}
\label{fig:volume}
\end{figure}

Figure~\ref{fig:volume} shows the daily volume of tweets by users classified as either pro- or anti-mask.
The two time series are highly correlated, with a Pearson correlation of $r = 0.938$.
The engagement begins roughly at the time of the first CDC recommendation to wear masks on 2020-04-03.
The increase in volume confirms a previous study which finds increased ``appetite'' to share opinions after major mask-related news \cite{cotfas2021unmasking}.
The subsequent peaks often revolve around major news stories involving masks, such as one when the U.S. Vice-President Pence visited a hospital without wearing a mask towards the end of April,\footnote{https://www.washingtonpost.com/politics/pence-meets-with-mayo-clinic-patients-staff-while-not-wearing-face-mask/2020/04/28/57c4200c-897e-11ea-9dfd-990f9dcc71fc\_story.html} continuous comparisons of the masking behavior of the two contenders for the U.S. Presidency in late May,\footnote{https://edition.cnn.com/2020/05/26/opinions/biden-mask-trump-ghitis/index.html} and subsequent adjustments to the guidelines by the public health officials who were trying to ``correct'' their previous messaging in mid-July.\footnote{https://edition.cnn.com/2020/07/12/politics/jerome-adams-surgeon-general-mask-mandate/index.html}
Given the high correlation of the two time series, we ask whether this effect is endogenous, i.e., if there is a causal feedback loop whereby one of the two stances answers the other.
A Granger causality test on these two time series for time lags from 1 to 14 days finds no strong relationship in either direction.
The likely implication of this negative result is that the volume of both stances has a common cause that is exogenous: external events that get discussed on Twitter.
Although we cannot exclude an effect with lag shorter than one day, the fact that the networks of two stances are well separated makes this hypothesis less likely.

\begin{figure}[t]
\centering
\includegraphics[width=0.75\linewidth]{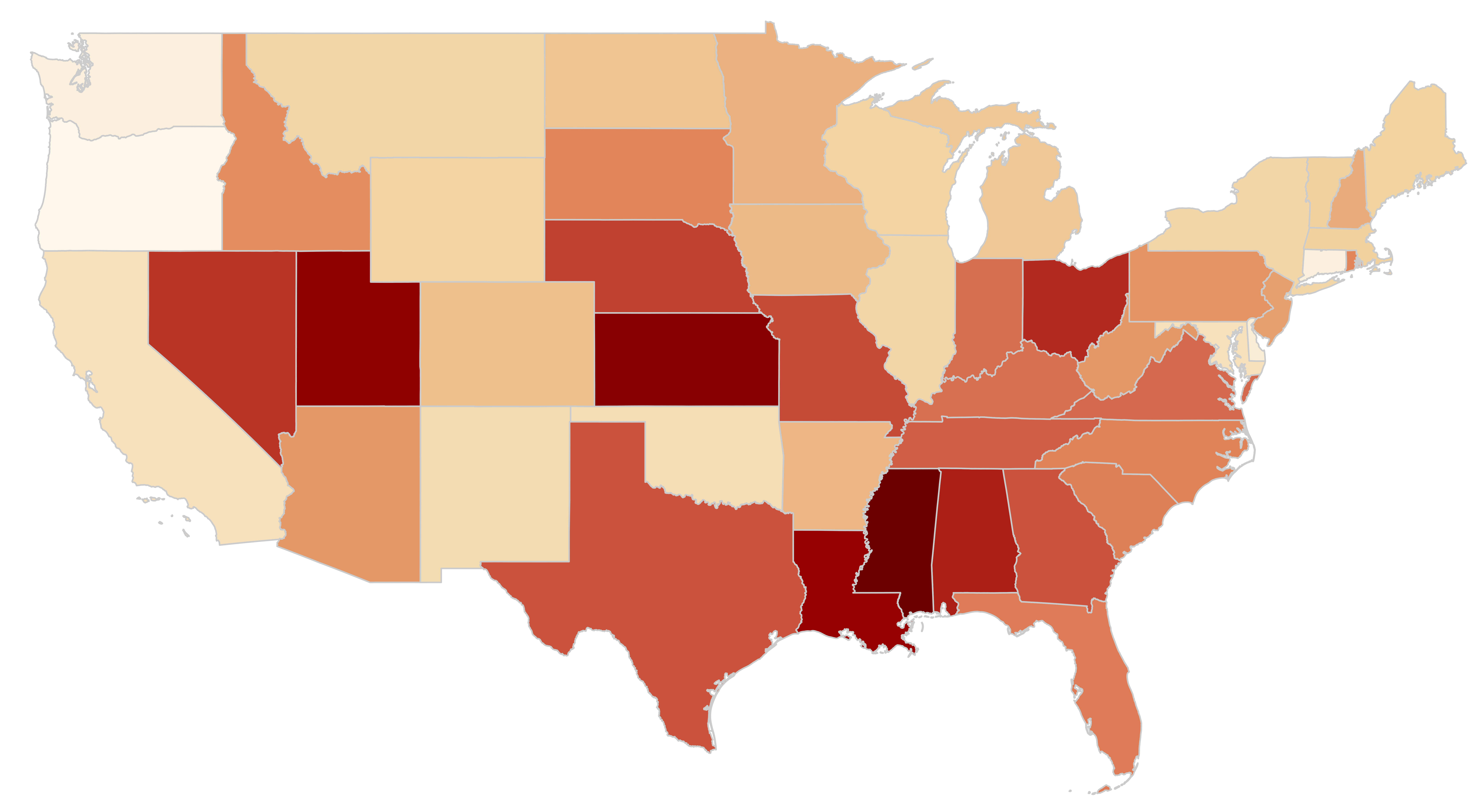}\hspace{0.5cm}\includegraphics[width=0.085\linewidth]{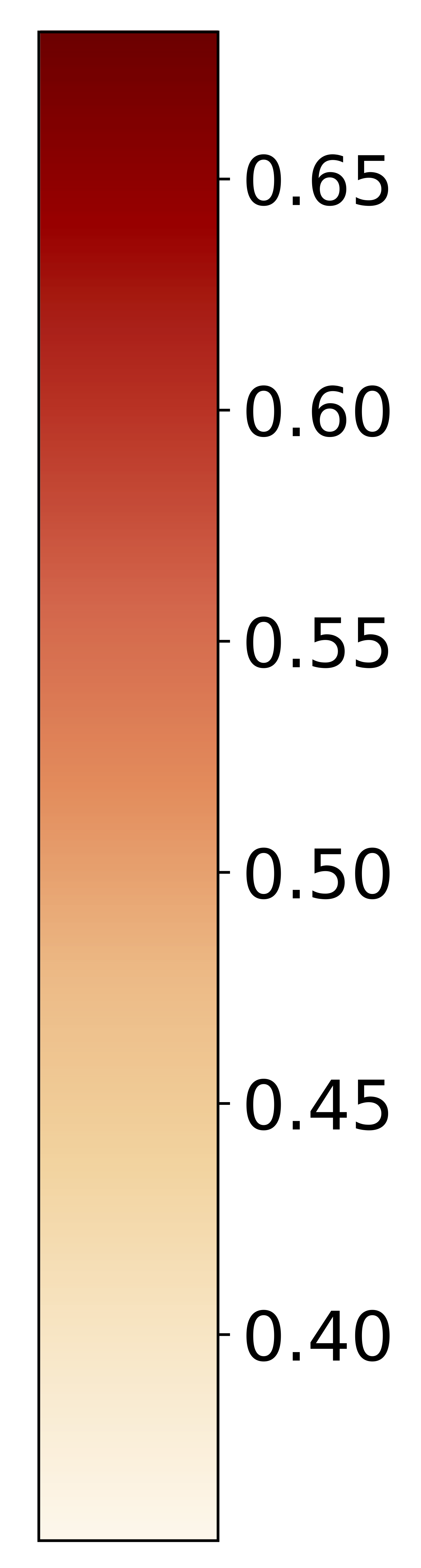}
\caption{Fraction of anti-mask users over the total geolocated users in the 48 U.S. states.}\label{fig:antimaskers_distribution}
\end{figure}

Figure~\ref{fig:antimaskers_distribution} shows the fraction of anti-mask users in each state. We find a particularly high concentration in the south, as well as some western states. These findings largely agree with contemporaneous studies of mask-wearing around the U.S.~\cite{katz2020map}.

\subsubsection{Political Leaning.}
As outlined in the Introduction, masking regulations have been strongly politicized. 
According to PEW, when surveyed in 2020, Republicans were more likely to call masks unnecessary, ineffective, oppressive, or unfair and were much less likely than Democrats to be concerned about unknowingly spreading COVID-19 to others.\footnote{\url{https://www.pewresearch.org/fact-tank/2020/10/29/both-republicans-and-democrats-cite-masks-as-a-negative-effect-of-covid-19-but-for-very-different-reasons/}}   

We compare the mask stance of the users we are able to classify to their political affiliation, which can be glimpsed via their Twitter social network. 
In particular, we consider the Twitter accounts with known political leaning that the users follows.
In accordance to previous work, we assume that users mostly follow accounts that are in agreement with their political views \cite{golbeck2011computing}, and create a list of prominent political accounts in order to propagate their leaning to their followers.
The list, which we make available to the research community,\footnote{Available at https://tinyurl.com/poliaccounts} includes 501 accounts of members of the U.S. Congress, 79 governors, 70 party entities, and 67 Attorney Generals, as well as 157 media accounts from allsides.com,\footnote{\url{https://www.allsides.com/media-bias/media-bias-ratings}} and 67 journalists from politico.com.\footnote{\url{https://www.politico.com/blogs/media/2015/04/twitters-most-influential-political-journalists-205510}}
In total, it has a rather balanced set of 487 right-leaning and 454 left-leaning accounts.
Note that we do not use prominent politicians such as the U.S. President, as such accounts may be followed purely out of their popularity or importance.
For each user in our mask-related dataset, we count the number of accounts they follow in our list: right-leaning $N_R$ and left-leaning $N_L$.
We consider only users who follow at least 5 accounts in our list from either side, and calculate the aggregated political leaning score as $S_{PL} = (N_R - N_L) / (N_R + N_L)$, which results in $S_{PL} \in [-1,1]$ with $1$ the most right-leaning score.
Thus, we are able to identify the political leaning of \num{18422} users.

Figure~\ref{fig:metis_hist_politics} presents the distribution of political leaning scores for the two categories of users based on their stances on masks.
Pro-mask users are more likely to be following left-leaning accounts, and anti-mask ones the right-leaning ones, with almost no users existing in the middle political ground.
In fact, we find a strong polarization at the extremes of the political spectrum, especially for anti-mask users.

\subsubsection{Moral Values.} Moral values are directly linked to decision making~\cite{karandikar2019predicting}. Since there is an altruistic component to mask wearing, we ask what are the moral values expressed by the holders of the two stances.
We assess the moral narratives by employing the MoralStrength lexicon~\cite{araque2020moralstrength}, which holds the state-of-the-art performance in moral text prediction.
MoralStrength lexicon provides, along with each lemma, the \textit{Moral Valence score}, a numeric assessment that indicates both the polarity and the intensity of the lemma in each moral foundation.
According to this lexicon, the Moral Valence is expressed in a Likert-scale from 1 to 9, with 5 to be considered as neutral.
Scores lower than 5 reflect notions closer to Harm, Cheating, Betrayal, Subversion, and Degradation, while values higher than 5 indicate Care, Fairness, Loyalty, Authority, and Purity.
For each lemma in a tweet and for each foundation, we obtain a moral valence score which is then averaged for each tweet.
Negation correction was not applied, as foundation polarities do not directly translate as opposites (e.g., “not care” is not the same as “harm”).
The MoralStrength lexicon has a limited linguistic coverage; as a result only the 41.5\% of the tweets were found to express a moral foundation. For all the rest, we assigned the value 5, neutral point of the Likert scale.
This approach pushes the observed mean towards the center of the scale, but captures the variability of the value across all documents (we discuss the implications of this methodological step in Discussion).
To assess statistical significance, we use a Student's t-test (with the Benjamini-Hochberg correction for multiple hypothesis testing) on the scores obtained from tweets written before and after the mandate date (2020-04-03) for each moral dimension.

\begin{figure}[t]
\centering
\includegraphics[width=0.85\linewidth]{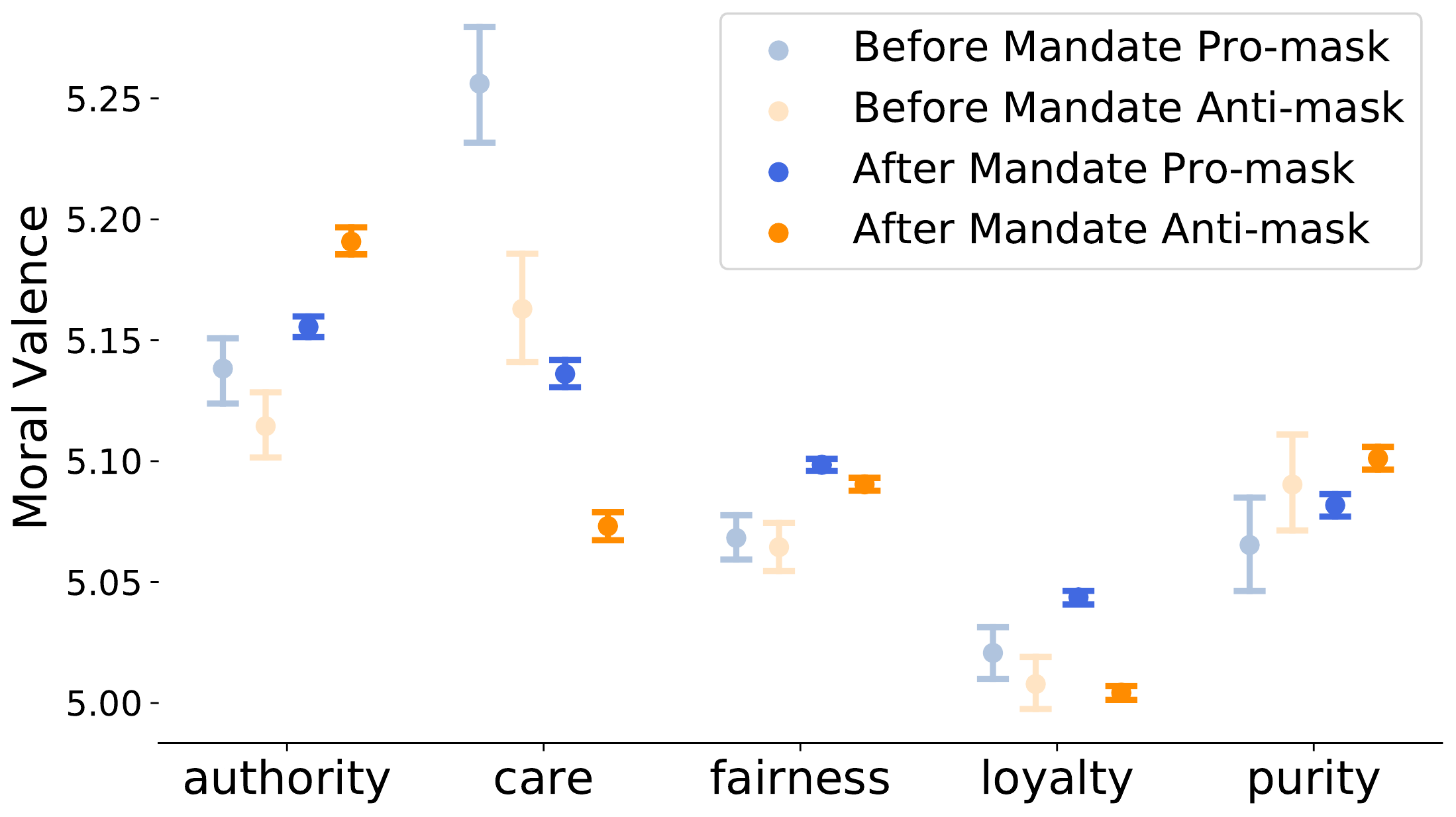}

\caption{Moral valence in narratives expressed by pro-mask and anti-mask users in the periods before (lighter points) and after (darker points) the mandate.
Dot represents the median value while the whiskers represent 5-95\% quantiles.}
\label{fig:morals_proanti}
\end{figure}

Figure~\ref{fig:morals_proanti} shows the mean moral value scores of each side in the periods before and after the mandate.
Before, the two sides display comparably similar values, except for \emph{care}, which is by far higher for the pro-mask side (significant at $p < 0.001$).
However, there is a clear shift in the moral narratives expressed after the mandate by both sides of the debate. 
To understand the context of these morally-charged expressions, we examine a sample of tweets for each side and value. 

First, we find an increase in the valence of \emph{authority} for the anti-mask side ($p<0.001$), which is mostly accompanied by criticism and mistrust of the decisions made by the authorities.
For instance, the anti-mask side associates wearing a mask with weakness and lack of leadership (\textit{``@JoeBiden Real leadership? With that thing on you look feeble''}). 
Conversely, the pro-mask side sees a lack of leadership in former President Trump's refusal to wear a mask (\textit{``Real leaders won't mind when the mask smears your absurd orange makeup''}). 
Most of the examples we find in this moral category are indeed criticism of the authorities, which nevertheless signifies that authority is held in high importance, especially for the anti-mask side (it is the value with the highest valence).
The fact that post-mandate the authority-related keywords have higher valence on the anti-mask side suggests stronger criticism of the authorities than the pro-mask side (for whom the increase is significant only at $p=0.004$ before the correction).

In terms of \emph{care}, both sides have a downwards shift after the mandate.
For the pro-mask side, this shift is accompanied by an increase in \emph{fairness} and \emph{loyalty}, which can be interpreted as a shift in focus from personal choice based on caring for others to complying with the mandate.
Also, in this case, the spotlight is often on the opposite side (\textit{``Masks protect others, which Trump doesn't care about. He cares only about himself.''}). 
Conversely, anti-mask supporters express themselves by prioritizing much less the notion of \emph{care}, explicitly showing disregard for the protection of others, or simply stating that they do not care about being criticized for not wearing a mask 
(\textit{``I'm tired of being called a murderer because I don't wear a mask. Cannot understand how people are such thoughtless idiots''}). 

In addition, pro-mask supporters express significantly more \emph{loyalty} in their messaging after the mandate ($p<0.0001$).
Upon examining a sample of posts, we find the increase is primarily related to criticism of those not wearing masks as loyalists to a political affiliation or to Trump personally (\textit{``People don't wear masks because of Trump. Dear leader doesn't wear one, loyal followers go along''}). 

After the mandate, the \emph{fairness} value increases for both sides (both at $p<0.0001$).
Narratives about the fair treatment of individuals appear to be present equally on both sides, focusing on negotiation when wearing masks is reasonable (pro-mask: \textit{``What if a person not wearing it understands the risks?''}), 
comparison with other rights violations (pro-mask: \textit{``The only ``rights'' that are violated is just not wearing a mask? They don't appreciate the nice life they have.''}), 
or whether the criticism is fair (anti-mask: \textit{``Colorado Governor Says People Who Refuse to Wear Mask are ``Selfish Bastards'' OK, and anybody who wants others to harm their immune system is a POS Marxist propagandist''}). 
Instead, the valence of \emph{purity} does not change substantially.
We find that the notion of purity is not tied to religious views, but to what people consider `natural' and `healthy' (e.g. \textit{``My 8 y.o. niece will be heading back to school but will have to wear a mask! How is it physically or emotionally healthy?''}). 



\begin{figure}[t]
\centering
    \includegraphics[width=0.90\linewidth]{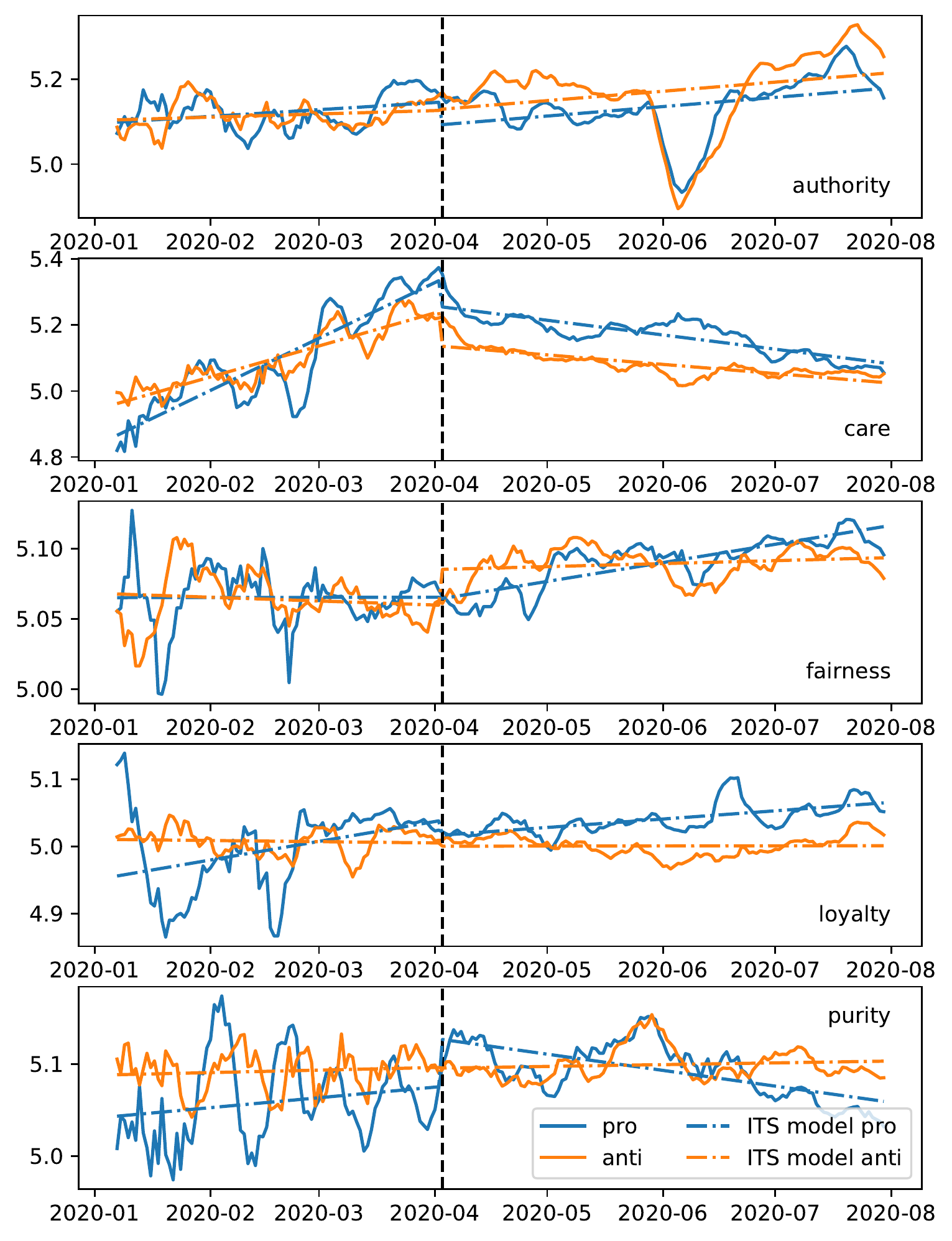}
    \caption{Time series of moral value scores of pro-mask and anti-mask users, along with an interrupted time series analysis model.}
\label{fig:moral_its}
\end{figure}

We further examine the evolution of moral values expressed by the two sides in time and model it via an Interrupted Time Series (ITS) linear model. 
Figure~\ref{fig:moral_its} depicts the evolution of the average moral foundation score per side, with a vertical dashed black line indicating the date of the official mask mandate.
Complementing our previous analysis, the interrupted time series model shows that for all the moral dimensions, after the mandate, there is an evident change in behavior by both sides. 
Perhaps the most interesting moral dimension is \emph{loyalty}, whose signal is evidently diverging for two sides exactly after the mandate date and continues the same trend until the end of our data collection. 
We also observe that not only does the value of \emph{care} decreases, the trend is downward over time, signaling a progressive shift in the debate.
Similarly, the value of \emph{purity} has a progressively negative trend for pro-mask side over time.
Thus, we find that the temporal dimension of the data can be instructive about the evolution of the rhetoric in terms of divergence between the two sides of conversation and changes in emphasis.

\subsubsection{Collectivism vs Individualism.}
\begin{table}[t]
    \caption{Use of singular and plural personal pronouns in a tweet by side, before and after the mask mandate. } 
    \label{tab:i_we}
    \centering
    \begin{tabular}{l c c c c}
    \toprule
    Pronoun        &  \multicolumn{2}{c}{Singular} & \multicolumn{2}{c}{Plural} \\
            \cmidrule(lr){2-3} \cmidrule(lr){4-5}
    Mandate	&  before & after  & before & after \\
    	\midrule
Pro-mask &  0.55 &  0.48 &  0.11 &  0.10 \\
Anti-mask & 0.53 &  0.60 &  0.09 &  0.10 \\
    \bottomrule
\end{tabular}
\end{table}

One of the main purposes of mask wearing is the protection of others, an expression of solidarity within the in-group against an external threat.
Thus, we turn to the Individualism-Collectivism (IC) dimension \cite{hofstede2001culture}, which captures the standing of individuals as interdependent members of a collective. 
We operationalize it via the personal pronouns used in the tweets, mainly first-person singular (``I'', ``me'',``mine'' etc.) and first-person plural (``we'', ``us'', ``ours'' etc), following existing literature~\cite{twenge2013changes}.

Table~\ref{tab:i_we} shows the average usage of the two sets of pronouns in the tweets posted by the two sides of the debate, separately before and after the mask mandate. 
In the period before mask mandate, anti-mask supporters use \emph{I} and other singular pronouns at 0.53, and after the prevalence increases to 0.60 ($+0.07$, $p<0.0001$), which points to an increased focus on the individual. 
Instead, pro-mask supporters decrease their usage of singular pronouns ($-0.07$, $p<0.0001$).
The mention of \emph{We} and other plural pronouns by both pro- and anti-mask supporters remain at the same levels ($\pm 0.01$), thus indicating no significant shifts in the focus on the self-identified group.
Thus, although having comparatively similar usages of singular pronouns before the mandate, the debate after the government's messaging becomes more individualistic for anti-mask side and less so for advocates of masking.

\subsubsection{Information Environment.} 


Finally, we turn to the information context where the debate takes place, in form of links to external sources and information from peers.
We begin by performing basic normalization steps on the original tweet text, including removing punctuation, accents, contractions, and stopwords, substituting emojis with a text description,\footnote{Using \texttt{emoji} library \url{https://pypi.org/project/emoji/}.} and finally by performing lemmatization.
We then identify the most distinguishing words used by each side by calculating the odds ratio of using particular words (lemmas) by one side compared to the other before and after the mask mandate.
We filter these words by their total frequency since otherwise rare terms would emerge as most distinguishing.  
Because the periods before and after the mandate have significant differences in volumes, we set the word frequency threshold to $100$ and $800$ for the two periods, respectively.
Politician names appeared in the top terms and were grouped together, hence resulting in two-word terms. 
The resulting top terms are:
\begin{itemize}
    \item Pro-mask: \emph{trumpvirus, Louie Gohmert, Putin, Herman Cain, wearadamnmask, penny, Mayo, GOP, DeSantis, covidiots}
    \item Anti-mask: \emph{riot, micron, virtue, leftish, loot, sheeple, unhealthy, bacterium, MSM, antifa}
\end{itemize}

The most distinctive words by each side are highly politicized: aside from references to politicians (similar to other recent findings around masking \cite{sanders2021unmasking}), we find several words used to attack the other side.
For instance, pro-maskers often use the term `trumpvirus' to refer to COVID-19, 
as a political response against the term `chinavirus' used by President Trump: 
\textit{``$@$realDonaldTrump  \#HermanCainRIP one more death due to \#TrumpVirus . Just think. If only he wore a mask and NOT attended the Tulsa \#CoronavirusRally''}. 
The focus on individuals is worth of notice, as in the case of Luie Gohmert, who strongly supported the use of hydroxychloroquine and attended a House Judiciary Committee hearing without wearing a mask, or in the case of Herman Cain, a Republican politician who opposed the mask mandate and later died of COVID-19.\footnote{\url{https://en.wikipedia.org/wiki/Herman_Cain#Health_and_death}}
Similarly, the derogative term `covidiots' is used to describe anti-mask supporters.

The opposite side instead focuses on the `riots' that would happen if lockdown and mask mandates are enforced, and on the claim that masks can stop a `bacterium' but not a `micron'-size virus. 
There are also claims that masks are `unhealthy' as they impede breathing, and are instead just `virtue signaling' devices: 
\textit{``Virtue signaling: mask and gloves. People lived with coronaviruses for 100+ years. Turn off the TV.''} 
A right-wing, anti-establishment sentiment can be inferred from the reference to `sheeple': \textit{``And the CDC telling everyone to mask up is just another test to see how long the sheeple will obey. \#Globalists''}, 
and from references to supposedly derogative terms such as `leftish' and `antifa'.
Nonetheless, what is labeled as the anti-mask side is also more varied in its opinions, ranging from just hesitant to fully conspiratorial, somewhat similar to the spread of opinions around vaccines~\cite{cossard2020falling}.


\begin{table}[t]
    \centering
    \caption{LDA topics with the most representative words, extracted separately from tweets for each mask debate side. The topic prevalence is reported in parenthesis.}
    \label{tab:lda_topics}
    \footnotesize
    \begin{tabular}{p{0.01\linewidth}p{0.05\linewidth}p{0.75\linewidth}}
    \toprule
    &Topic & \multicolumn{1}{c}{Words} \\
    \midrule
    \parbox[t]{0mm}{\multirow{6}{*}{\rotatebox[origin=c]{90}{pro\hspace{-2mm}}}}  
    &T1 (.35) & distance, wear, social, day, mandate, today, friend, hope, state, stay\\
    & T2 (.33)&  wear, people, virus, covid, bad, protect, medical, infect, spread, science\\
    &T3 (.31) & wear, Trump, Cain, rally, covid, Herman, die, Tulsa, death, kill\\
    \midrule
    \parbox[t]{0mm}{\multirow{6}{*}{\rotatebox[origin=c]{90}{anti\hspace{-2mm}}}}     
    & T1 (.46) & wear, people, covid, virus, work, distance, die, thing, protect, spread\\
    & T2 (.31) & wear, mandate, people, state, store, vote, business, leave, today, order\\
    & T3 (.22) & man, Trump, rally, school, eye, fuck, kid, hope, stupid, big\\
    \bottomrule
    \end{tabular}
\end{table}


Moving to higher-level constructs, we aim to uncover common patterns in the argumentation proposed by both sides by applying a topic modeling approach base on  latent Dirichlet allocation (LDA)~\cite{blei2003latent}.
Limitations of LDA clustering of short text are known, still, it offers a good compromise between clustering performance and computational cost~\cite{qiang2020short}.\footnote{Using the Gensim library~\url{https://radimrehurek.com/gensim}}
To derive the optimum number of topics $k$, we optimize the topic coherency ($C_v$ metric~\cite{roder2015exploring}) models with $k \in [2,10]$.
For both the pro-mask and anti-mask sides, the model with $k=3$ is the best fit.
Table~\ref{tab:lda_topics} presents the salient keywords that form the corresponding topics.
The topics are ranked according to their prevalence, with T1 the most prevalent one, and similarly, the terms are ranked by descending importance for the specific topic. 

From the emergent topics, the most prominent one on the pro-mask side concerns the various interventions, including \emph{social distancing} and \emph{wearing} a mask.
(\textit{``COVID is not a flu! Everyone needs to wear a mask to protect others from these germs.''}), 
echoing our earlier finding of higher care value of this side.
The second topic includes references to \emph{medicine} and \emph{science}, while the third centers around (Republican) political figures.
On the anti-mask side, the most prominent topic also concerns the interventions, but instead focuses on whether interventions \emph{work} against the \emph{spread}.
The second one puts the mandate in the context of the \emph{businesses} and \emph{stores}, in contrast with T1 from the pro-mask side which speaks about other interventions such as social \emph{distancing} and \emph{staying} at home.
The third topic is about political \emph{rallies}, and argues that masks are not useful (since the \emph{eyes} are exposed to the virus: \textit{``Still, even with a mask, you are uncovering the mucous membranes in your eyes''}). 

\begin{table}[t]
    \centering
    \caption{Counts of the top 30 URL domains posted by pro- and anti-mask users. Domains colored by class: news and news aggregators (black), social media and social media automator/aggregators (\textcolor{BrickRed}{red}), business platforms (\textcolor{Blue}{blue}), medical organization (\textcolor{OliveGreen}{green}).}
    \label{tab:top_urls}
    \footnotesize
    \begin{tabular}{r l r l}
    
\multicolumn{2}{c}{Pro-mask } & \multicolumn{2}{c}{Anti-mask}   \\
\textcolor{Black}{rawstory.com} & 2317 & \textcolor{BrickRed}{youtube.com} & 3485 \\
\textcolor{Black}{cnn.com} & 2000 & \textcolor{Black}{thegatewaypundit.com} & 1341 \\
\textcolor{BrickRed}{youtube.com} & 1751 & \textcolor{Blue}{etsy.me} & 1210 \\
\textcolor{Black}{washingtonpost.com} & 1393 & \textcolor{BrickRed}{instagram.com} & 912 \\
\textcolor{Black}{a.msn.com} & 935 & \textcolor{Black}{foxnews.com} & 903 \\
\textcolor{Black}{apple.news} & 872 & \textcolor{Blue}{zazzle.com} & 796 \\
\textcolor{Black}{huffpost.com} & 758 & \textcolor{Black}{breitbart.com} & 781 \\
\textcolor{Black}{news.yahoo.com} & 686 & \textcolor{Black}{nypost.com} & 472 \\
\textcolor{Black}{flip.it} & 630 & \textcolor{Blue}{fineartamerica.com} & 453 \\
\textcolor{Black}{nytimes.com} & 587 & \textcolor{Black}{fxn.ws} & 393 \\
\textcolor{Black}{nbcnews.com} & 573 & \textcolor{Black}{westernjournal.com} & 362 \\
\textcolor{Black}{thehill.com} & 527 & \textcolor{BrickRed}{dlvr.it} & 344 \\
\textcolor{Black}{dailykos.com} & 486 & \textcolor{Blue}{pixels.com} & 317 \\
\textcolor{BrickRed}{instagram.com} & 458 & \textcolor{BrickRed}{buff.ly} & 288 \\
\textcolor{Black}{businessinsider.com} & 449 & \textcolor{Black}{theblaze.com} & 282 \\
\textcolor{Black}{thedailybeast.com} & 402 & \textcolor{Black}{bizpacreview.com} & 268 \\
\textcolor{Black}{newsweek.com} & 371 & \textcolor{Black}{infowars.com} & 250 \\
\textcolor{Black}{theguardian.com} & 343 & \textcolor{Blue}{etsy.com} & 246 \\
\textcolor{Black}{usatoday.com} & 337 & \textcolor{BrickRed}{ift.tt} & 236 \\
\textcolor{Black}{yahoo.com} & 333 & \textcolor{Black}{cnn.com} & 231 \\
\textcolor{Black}{cnbc.com} & 307 & \textcolor{Black}{twitchy.com} & 217 \\
\textcolor{Black}{politico.com} & 301 & \textcolor{Blue}{ebay.us} & 202 \\
\textcolor{Black}{newsbreakapp.com} & 295 & \textcolor{OliveGreen}{ncbi.nlm.nih.gov} & 196 \\
\textcolor{BrickRed}{buff.ly} & 260 & \textcolor{BrickRed}{facebook.com} & 192 \\
\textcolor{Black}{npr.org} & 252 & \textcolor{OliveGreen}{nejm.org} & 191 \\
\textcolor{Black}{politicususa.com} & 236 & \textcolor{BrickRed}{newsbreakapp.com} & 175 \\
\textcolor{Black}{apnews.com} & 231 & \textcolor{Black}{a.msn.com} & 173 \\
\textcolor{Black}{nypost.com} & 223 & \textcolor{Black}{google.com} & 162 \\
\textcolor{Black}{latimes.com} & 209 & \textcolor{Black}{dennismichaellynch.com} & 152 \\
\textcolor{Black}{mol.im} & 208 & \textcolor{OliveGreen}{aapsonline.org} & 142\\
    \end{tabular}
\end{table}


Finally, the two sides of the debate have about the same proportion of tweets with URLs (around 23\%).
Earlier works have shown a polarization in terms of news sources accompanying political polarization~\cite{garimella2021political}, however here we find differences in behavior beyond sharing news media.
Table~\ref{tab:top_urls} shows the top domains of the URLs posted by pro- and anti-mask users, along with the counts.
Pro-mask users overwhelmingly post URLs pointing to news websites or aggregators.
YouTube and Instagram feature prominently in both lists, though anti-mask users favor YouTube more than twice the second most popular domain.
Anti-mask tweets also link to a variety of business platforms, including Etsy and Ebay, and lesser-known ones such as Zazzle, a platform for custom-designed products.
In addition, anti-mask users link to the governmental agency National Center for Biotechnology Information (NCBI), the New England Journal of Medicine (NEJM), and the Association of American Physicians and Surgeons (AAPS).
These findings stand in contrast to a smaller recent study of geolocated-only tweets~\cite{he2021people} that finds that anti-mask tweets were less likely to share external information from public health authorities.
An explicit comparison of the captured tweet sets would be necessary to resolve these findings.

\section{Discussion \& Conclusions}


Our analysis reveals that the government messaging about mask wearing provoked---instead of the intended focus on the benefits to the communities and society at large---a marked shift in the moral values towards higher politicization of the issue, with an increased focus on authority and fairness.
We argue that interventions targeted to those resistant to mask-wearing should center around these values, instead of appealing to those valued less as the debate goes on.

It is no surprise that we find the mask debate captured on Twitter to be highly politically polarized. 
It is well-known that polarization in the U.S. political scene has been growing in the last decades,\footnote{\url{https://www.pewresearch.org/politics/2021/11/09/beyond-red-vs-blue-the-political-typology-2/}}   
and this growing polarization has important effects on several areas, including public health.
As polarization and health behaviors intersect, it is crucial to understand their interaction to design effective health policies.
For instance, the ideological and moral facets of health intervention perceptions are closely related to the compliance in the population, and the resulting effectiveness, as shown for instance in recent statistics relating vaccination uptake to political leaning.\footnote{\url{https://acasignups.net/21/07/20/update-us-covid19-vaccination-levels-county}}  
However, shifting established associations between health-related beliefs and political leaning may be challenging. 
For instance, the attendees of Donald Trump's rally in August 2021 booed him after he voiced support for the COVID-19 vaccination drive,\footnote{\url{https://www.theguardian.com/us-news/2021/aug/22/donald-trump-rally-alabama-covid-vaccine}} thus illustrating that even the forerunners of the Republican party may encounter challenges in connecting with their constituents.
As the pandemic develops, and more mandates are issued by the governments, constant polling and monitoring is essential in establishing the public response to these measures (as of mid-2021, the attitudes toward masking are still highly polarized).\footnote{\url{https://www.politico.com/news/2021/08/02/poll-americans-back-return-of-masking-502144}}


We find that different moral values underpin the reasoning emphasized by the two camps.
Pro-mask arguments highlight loyalty and fairness while criticizing the opposing leadership.
The anti-mask ones, on the other hand, focus on the authoritarian and oppressive aspects of the mandate and show a lack of concern for the effects of their actions on others.
In accordance with these results, when examining this phenomenon through the lenses of the Collectivism-Individualism theory, we notice a decisive shift of the anti-mask community towards individualism, with more intense use of first-person personal pronouns.
We note the lack of loyalty among the values emphasized by the anti-mask side, which tends to hold a conservative political view, and differs from the commonly observed ones associated with conservatism: authority, loyalty, and purity~\cite{kivikangas2021moral,fulgoni2016empirical}. 
This may be a response to the pro-social framing of the mask intervention, thus leading its opponents to de-emphasize the in-group narrative usually common to their side.
This interpretation may point to motivated reasoning, wherein the desired conclusion modifies the worldview usually taken.
Interestingly, the ITS analysis shows divergence over time between the two groups on the value of loyalty, which is alarming since such disagreement, if not adequately addressed, can lead to severe societal polarization.


This polarization seems to have been accompanied by a lively commercial activity. 
When examining the links posted by the two sides of the issue, we find a prominent existence of commerce-related platforms including \emph{etsy.com}, \emph{ebay.us}, as well as platforms for custom creation of merchandise such as \emph{zazzle.com}. 
Indeed, a brief search on these websites reveals an assortment of t-shirts, coffee mugs, face masks, and baby bibs with political messages from both sides of the debate. 
Sometimes historical symbols were used to make a stance, such as the use of the yellow star---like those forced on Jews by Nazi Germany---which was sold by a Nashville store protesting against the vaccination campaign.
After community criticism, the item was removed\footnote{\url{https://www.bbc.com/news/world-us-canada-57297902}}.  
Our findings suggest that there is an active development of symbolism and aesthetics of the resistance movement, and it would be a fascinating subject of research to uncover the non-verbal representations of the self and the group, expressions of values, and calls to action \cite{mcgarry2019aesthetics,awad2020protest}.
Awareness of such symbolism and self-conceptualization is vital for crafting appropriate messages and fostering communication between the two sides.


Despite the volume and span of the data analyzed here, this study has some limitations as most social media studies. 
Especially when concerning politically charged topics, often a minority of vocal users dominates the conversation~\cite{Mustafaraj2011}, thus making the opinions of the ``silent majority'' challenging to discern.
The observed relationships between the stances on masking, moral values, and political stance are limited to those who choose to express themselves vocally.
In addition, we use a high-precision vocabulary to measure moral values, which has the side effect of pushing the average valence in our results close to the neutral point, due to a large number of tweets for which we cannot extract moral valence.
Traditional surveys are necessary to reach those not as comfortable expressing their opinions online; however, even those have selection biases.
The automated tools utilized in this study are not perfect: after manual examination, we find that the network-based classifier achieved an accuracy of 72.4\% for the anti-mask class, thus introducing noise in the subsequent analysis.
However, from the experience of manually labeling the users, we postulate that the complexities of human expression, including humor and sarcasm, may limit the best possible performance of such a classifier.
Similarly, the inference of moral values from the text may struggle with vocabulary mismatch, sarcasm, and self-censorship. 
Finally, the generalizability of this study is limited by the unique circumstances of an unprecedented global pandemic happening in a hyper-connected world during a politically polarized social environment.
These considerations must be taken into account when comparing our findings to new scenarios.

\subsection{Ethical Considerations}

The dataset presented here contains only tweets which were publicly available at the time of the collection.
We make the dataset publicly available to the research community in compliance with the Twitter Terms of Service,\footnote{\url{https://developer.twitter.com/en/developer-terms/}} that is, sharing only the tweet IDs of the collected posts, which will have to be re-collected. 
In this paper, we have rephrased all quoted tweets to prevent re-identification of their authors.
This practice ensures that the tweets which have been removed (either by the user or the platform) will not be available.
Although large, this dataset does not include users with particular disabilities which may disallow them to interact with the platform, as well as minors and those blocked by Twitter.
On the other hand, the content collected here affects not only those who have posted it, but also those who viewed or interacted with it, which may be orders of magnitude more users, since most users are ``lurkers'' who consume social media content without posting \cite{van20141onepercent}.
Ultimately, the masking decisions made by people engaging in this deliberation may directly affect the health and life of vulnerable people, such as those with autoimmune disorders or other conditions making them especially vulnerable to COVID-19. 
Additionally, the sometimes aggressive rhetoric in this material may not be suitable for young Twitter users, or those dealing with mental health issues. 
Also, although the focus of this study is the moral dimension of the debate, we caution public health communicators not to overemphasize moral or emotional dimensions of their message (or attempt to emotionally manipulate their audience), but rather provide the clearest and most informative messaging possible.
Further, we would like to discourage the tools used in this study to be used for targeting individuals espousing particular opinions for harassment or undue surveillance, and to follow the AAAI ethical guidelines\footnote{\url{https://www.aaai.org/Conferences/code-of-ethics-and-conduct.php}} in the application of these findings.

\end{document}